\documentclass[journal=nalefd,manuscript=article,layout=twocolumn]{achemso}

\usepackage[version=3]{mhchem} 

\author{Dmitry Kazantsev}
 \email{kaza@itep.ru}
 \affiliation[University of Erlangen]
{LEB, University of Erlangen, Cauerstr. 6, D-91058, Germany}
\altaffiliation{Institute for Theoretical and Experimental Physics,
Moscow (perm.pos.)}

\title[\texttt{achemso} demonstration]
{ASNOM mapping of SiC epi-layer doping profile and of surface phonon
polariton waveguiding}

\begin{document}
\begin{abstract}
  The apertureless SNOM mapping of the slightly-doped 4H-SiC epitaxial
layer grown on a heavily-doped 4H-SiC substrate was performed with a
cleaved edge geometry. ASNOM images taken at the light frequencies
of a $C^{13}O_{2}^{16}$ laser show a clear contrast between the
substrate and the epitaxial layer. The contrast vanishes at the
laser frequency of $884cm^{-1}$, and gets clearer at higher
frequencies $(923cm^{-1})$.
This can be explained by changes in the local polarizability of SiC
caused by the carrier concentration, which are more pronounced at
higher frequencies. Since the light frequency is tuned up further
($935cm^{-1}$), a transversal mode structure appears in the ASNOM
map, indicating a waveguide-like confinement of a surface phonon
polariton wave inside the strip of an epi-layer outcrop.
\end{abstract}

\section{Introduction}
Recently, several papers report a successful application of an
Apertureless Scanning Near-field Optical Microscopy (ASNOM, s-SNOM)
\cite{s_SNOM_first, SNOM_1nm} to the mapping of a solid state
surface optical properties~\cite{Knoll_OC2000,
Knoll_IR_conductivity_APL2000, Keilmann_THz_ASNOM_FET_Mapping,
Keilmann_Polytype_NL06}. This method demonstrates not only the
ability to distinguish different materials by their ASNOM
responce\cite{Keilmann_Nature99, APL_085_005064,
Hillenbr_FisherGrid_JM01, Hillenbr_MaterialMapping_APL02}, but is
even sensitive to the fine variations in an ASNOM response caused by
a presence of the free carriers in the same
media\cite{AdvMat_Keilmann_Infineon_ASNOM_Transistor,
Keilmann_THz_ASNOM_FET_Mapping, Hillenbr_InP_Nanowires_NL10}.

\section{Experimental}
\begin{figure}
\resizebox{0.45\textwidth}{!}{\includegraphics{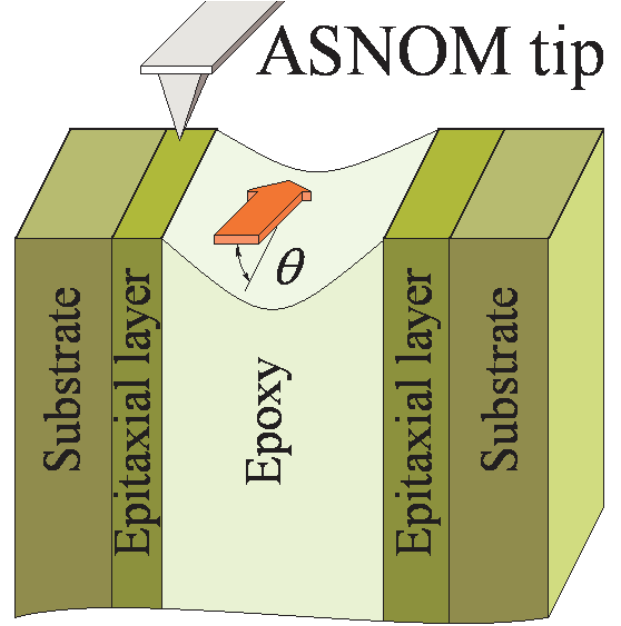}}
\caption{
\label{fig:Epi_SiC_Mapping_Config}
Geometry of the experiment. The slices of a SiC wafer with epitaxial
layer were glued together and then polished in order to get
mechanical and optical access to the cleaved edge. To make a
comparison of ASNOM response on two sides easier, the direction of
irradiating light (shown with an arrow) was chosen along the slit
between the wafer slices. The direction of the irradiating beam is
declined in respect to the surface plane ${\theta}=35^{o}$.
}
\end{figure}

In the present paper, we report an ASNOM mapping of a SiC epitaxial
layer (doping $n_N=9\cdot10^{14}cm^{-3}$, thickness $d=9.9{\mu}m$)
grown on a SiC substrate (doping $n_N=7\cdot10^{18}cm^{-3}$). Both
the substrate and the epitaxial layer are of 4H polytype, with the
c-axis normal to the surface (vicinal angle 4$^{o}$). The epitaxial
layer side of the wafer demonstrates a much stronger ASNOM response
than the substrate side. To prepare the samples for an experiment
(see Fig.~\ref{fig:Epi_SiC_Mapping_Config}), the wafer was cut into
slices of about 500$\mu$m in width, and the pairs of such slices
were glued onto a metal carrier piece, with their cut surfaces
facing up. After the epoxy got hardened, the sample was mechanically
polished in order to provide mechanical and optical access to the
media. Such a geometry prevents the edges of the wafer from cracking
during the polishing procedure, and also allows an easy a comparison
between the substrate and the epitaxial-layer edge response. In
addition, such a geometry prevents an ASNOM tip from being broken by
a sample edge.

A home-built system was used for ASNOM mapping of the sample. A
$C^{13}O_{2}^{16}$ laser was used as a light source to illuminate
the tip and to detect the scattered optical signal in a Michelson
interferometer scheme by optical
homodyning\cite{Keilmann_PTRS_362_787_Review,Taubner_NL04_Michelson}.
%
The light (of amplitude $E_{sc}(\vec{r}_{tip})$ corresponding to the
tip location $\vec{r}_{tip}$ with respect to a sample) scattered by
an ASNOM tip was collected with an objective and directed back to
the photodetector, overlapping the reference beam spot.
%
Since near-field optical interaction between the tip and the surface
depends on the tip-sample distance in a nonlinear way, the higher
harmonic components of a tip oscillation frequency $\Omega$ were
recovered\cite{Wurtz_ASNOM_RSI98,Knoll_OC2000} in the photocurrent
oscillations ${I_{det}(t)}$ as an ASNOM signal
${I_{det}^{(n\Omega)}}$. Averaging over the reference beam phase was
performed in order to exclude arbitrary origin of the reference arm
length. With this signal processing, an amplitude of
near-field-caused variations in $E_{sc}$ can be acquired (as a
complex number):
\begin{equation}
E_{sc}(\vec{r}_{tip}) {\propto}{I_{det}^{(n\Omega)}}
{\propto}E_{loc}(\vec{r}_{tip})\alpha_{eff}^{(n\Omega)}(\varepsilon_{s}(\vec{r}_{tip}),z_0)
 \label{eq:E_NF}
\end{equation}

where $\alpha_{eff}^{(n\Omega)}(\varepsilon_{s},z_0)$ is an
effective tip polarizability\cite{Hillenbr_FisherGrid_JM01},
depending on the surface local dielectric constant
$\varepsilon_{s}(\vec{r}_{tip})$ at the tip location. It also
depends on the tip vibration amplitude $z_{0}$ and its dielectric
constant.
%

\section{Results and discussion}
\begin{figure}
\resizebox{0.45\textwidth}{!}{\includegraphics{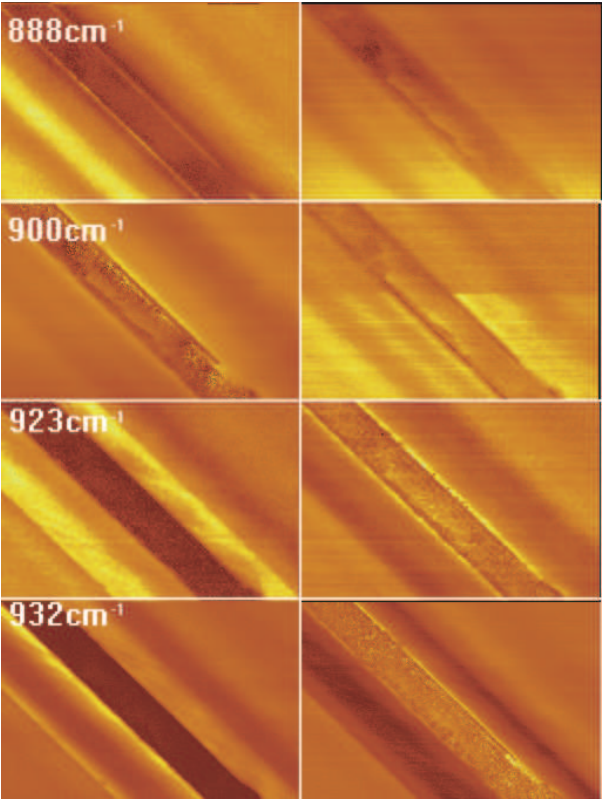}}
\caption{
\label{fig:Epi_SiC_ASNOM_maps}
 ASNOM maps of two SiC wafers, prepared in a cleaved-edge geometry.
The images were acquired at different light frequencies from
$888cm^{-1}$ to $935cm^{-1}$. Left maps represent ASNOM amplitude
$I_{det}^{(2\Omega)}$ collected at $2^{nd}$ harmonic of the tip
oscillation frequency $\Omega$, right ones contain the phase of
$I_{det}^{(2\Omega)}$ .
}
\end{figure}

The amplitude and phase maps of an ASNOM signal (oscillations in the
photocurrent recovered at the second harmonic of the tip oscillation
frequency, and then averaged over full turn of the the reference
phase) are shown in the Fig.~\ref{fig:Epi_SiC_ASNOM_maps}.

 Similar to results
reported in~\cite{Keilmann_Polytype_NL06}, ASNOM image taken at
$884cm^{-1}$ contains no step at all (and simultaneously a SPP wave
running from the sample edge is well seen, observed due to a large
lateral decay length), then (see image taken at $900cm^{-1}$) the
SPP wave gets weaker (but still no boundary is visible) and finally
a well-pronounced step appears at the frequency of $923cm^{-1}$,
between the substrate and the epitaxial layers.

Such steps can be well explained by the changes of
$\alpha_{eff}^{(n\Omega)}(\varepsilon_{s}(\vec{r}_{tip}),z_0)$ term
in (\ref{eq:E_NF}) caused by $\vec{r}_{tip}$ variations during the
sample scanning. A frequency-dependent dielectric function of SiC is
commonly ~\cite{Tiwald_PRB_60.11464_SiC_Ellypsometry,
harima_AP-1995_Raman, Sasaki_PRB40.1762_Raman,
Yugami_AP-1987.354_Raman} expressed as the sum of the Lorenzian term
(for the lattice) and the Drude term (for the free carrier
subsystem):
\begin{equation}
\varepsilon(\omega)=\varepsilon_{\infty}
(1+\frac{\omega^{2}_{L}+\omega^{2}_{T}}{\omega^{2}_{T}-\omega^{2}-i\omega\Gamma})
+\frac{\omega_p^2}{-\omega^{2}-i\omega\gamma}
 \label{eq:eps_omega}
\end{equation}
The values $\omega_{L}$ and $\omega_{T}$ denote experimentally
observed bulk phonon polariton frequencies, $\Gamma$ is the phonon
damping constant. Factor $\gamma$ in the Drude term denotes the
electron subsystem damping constant, and $\omega_{p}^2$ is the
plasma frequency
\begin{equation}
\omega_{p}^2= \frac{4{\pi}Ne^{2}}{m^*m_e}
 \label{eq:omega_el_}
\end{equation}
depending on the free carrier concentration $N$ and the effective
mass $m^*$. Strictly speaking, lattice and electron properties of
SiC are anisotropic, but for the 4H polytype this difference is just
a few percent\cite{Eps_SiC_PRB_1997}, so it can be neglected in the
first approximation. In our calculations, we used the following
values\cite{Eps_SiC_Landolt_B,
SiC_Properties_Bechstedt_PSSB-1997_p35}:
$\omega_{T}=797cm^{-1}$\cite{Sasaki_PRB40.1762_Raman},
$\omega_{L}=969cm^{-1}$ \cite{Sasaki_PRB40.1762_Raman},
$\Gamma=6cm^{-1}$\cite{harima_AP-1995_Raman},
$\varepsilon_{\infty}=6.7$. With an electron mass $m_e=0.4$ we
estimate free carrier plasma frequency $\omega_{p}$ as $90cm^{-1}$
for the epitaxial layer and $8000cm^{-1}$ for the substrate. Using
these values, $\alpha_{eff}^{(n\Omega)}(\varepsilon_{s},z_0)$ was
calculated. Its dependency on the local dielectric constant of the
surface, considered in long-wave dipole approximation
\cite{Hillenbr_FisherGrid_JM01} of the
Pt-coated\cite{Pt_epsilon_Ordal_AO-1985} tip response in the
vicinity of the sample already gives a good fit.

%
 In our case, however, the images acquired with an ASNOM
at the higher light frequency of (see $935cm^{-1}$ image in Fig.
\ref{fig:Epi_SiC_ASNOM_maps}) can not be interpreted as just changes
in the local value of the sample dielectric function (see term
$\alpha_{eff}^{(n\Omega)}(\varepsilon_{s},z_{0})$ in the
(\ref{eq:E_NF}) expression). Unlike the images presented
in~\cite{Keilmann_Polytype_NL06} we observe clear wave phenomena,
most likely due to less lateral decay of the SPP wave in our
samples. A sinusoid-like distribution of $I_{det}^{(2\Omega)}$
amplitude and phase across the epitaxial layer outcrop is clearly
seen. It can not be reasonably explained by any distribution of the
sample dielectric properties caused e.g. by the doping profile. In
this case, additionally to the calculation of a tip effective
polarizability in (\ref{eq:E_NF}) we have to take into account a
collective electromagnetic effect, namely an excitation of the
running SPP waves by the laser light over all points of the sample
surface. A focal spot of the objective is relatively large
($30-50{\mu}m$) so that the laser also irradiates some sample area
around the tip. The elementary electromagnetic excitations created
by a laser light in some other points of the surface, are then
delivered as SPP waves to the point of probing.
 Therefore, we cannot consider a local field amplitude
$E_{loc}(\vec{r}_{tip})$ in expression (\ref{eq:E_NF}) as a
constant\bibnote{In our setup, the tip is always located in the
focus of the objective, so that the amplitude and especially phase
of the laser light at the tip location are constant.}, but have to
write it as $E_{loc}(\vec{r}_{tip})=E_{las}+E_{spp}(\vec{r}_{tip})$
instead. Depending on the phase difference, the interference between
these two terms in the sum might be either constructive or (as takes
place in Fig.\ref{fig:Epi_SiC_ASNOM_maps}) destructive.

The homogeneous problem to describe an SPP wave on infinite surface
of a polar crystal was solved \cite{LST_1941, Born_book_1954,
Barron_1961} by substitution
\begin{equation}
E_{spp}(\vec{r}_{xyz},t)=E_{spp0}e^{i\vec{k}_{xy}(\omega)\vec{r}_{xy}}
e^{-\delta_{z(b,a)}z}e^{i{\omega}t}
\label{eq:E_spp}\\
\end{equation}\\
The problem eigenvalues were found for the lateral propagation term
\begin{equation}
k_{xy}(\omega)=\frac{\omega}{c}\sqrt{\frac{\varepsilon_{vac}\varepsilon_{SiC}
(\omega)} {\varepsilon_{vac}+\varepsilon_{SiC}(\omega)}},
\varepsilon_{vac}\equiv1,\label{eq:k_xy}\\
\end{equation}\\
and for the factor $\delta_{z(b,a)}$ describing exponential
amplitude decay beneath and above the surface, respectively. To our
knowledge, the \emph{inhomogeneous} task of describing SPP wave
excitation by an external wave is not solved rigorously yet, in the
general case of an arbitrary surface shape.

\begin{figure}
\resizebox{0.45\textwidth}{!}{\includegraphics{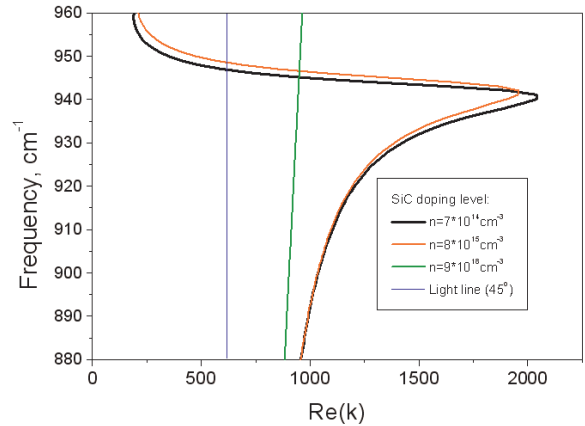}}
\caption{
\label{fig:DispersionLaw}
Surface phonon polariton dispersion law plotted for different values
of the free carrier concentration in SiC. Most differences occur at
the frequencies close to the LO side of the RSB, where the
denominator $\varepsilon(\omega)+\varepsilon_{vac}(\equiv1)$ in
Eq.\ref{eq:k_xy} crosses zero.
}
\end{figure}

A dispersion law of the SPP waves calculated with the expressions
mentioned above is shown in fig.~\ref{fig:DispersionLaw}. One can
see that the plot curves of different doping level are very close to
each other at the frequencies of $880-900cm^{-1}$. At the
frequencies of $920-940cm^{-1}$ the curves diverge dramatically. A
substrate ($n_N=7\cdot10^{18}cm^{-3}$) demonstrates more
\emph{plasmon}-polariton than \emph{phonon}-polariton behavior, so
that the Z-loop on the curve vanishes completely. Therefore, on a
cleaved sample side, a strip of undoped epi-layer outcrop appears to
be surrounded by the surface media of metal-like electromagnetic
properties, with a sharp step at the interface. In such a case, the
SPP wave excited on a SiC surface by the irradiating laser light
gets confined in a two-dimensional waveguide. As it can be seen in
the Fig.~\ref{fig:DispersionLaw}, the SPP wavelength at the
frequency of $935cm^{-1}$ gets significantly shorter than at the
frequency of $923cm^{-1}$. Consequently, a transversal mode field
distribution appears in the ASNOM map at $935cm^{-1}$, similar to
those known for the cm-band waveguides.

In conclusion, we studied ASNOM mapping of a doping profile in a
4H-SiC sample. Due to the differences in SiC dielectric function
caused mainly by the free carrier plasma in the substrate, we
observe clear doping contrast with a lateral resolution of 20nm. The
sharp step in the SPP dispersion law on the surface areas of
different doping leads, at some light frequencies, to a confinement
of the running SPP wave in the strip defined by the outcrop of a
low-doped epitaxial layer, similarly to the light confinement in
conventional optical fiber.

%

%




\acknowledgement

Surface maps were acquired and stored with a help of R9 controller
(RHK Tech. MA). The help of K.Kollin (native English speaker) is
acknowledged for his kind proof of the text. The work was supported
by the German National Science Foundation (DFG grant KA3105/1-1).

\bibliography{SiC_Epi-layer}

\begin{mcitethebibliography}{28}
\bibitem[Not()]{Note-1}
In our setup, the tip is always located in the focus of the objective, so that
  the amplitude and especially phase of the laser light at the tip location are
  constant.\relax
\mciteBstWouldAddEndPunctfalse
\mciteSetBstMidEndSepPunct{\mcitedefaultmidpunct}
{}{\mcitedefaultseppunct}\relax
\EndOfBibitem
\end{mcitethebibliography}

\end{document}